**Extraction of emission parameters for large-area field emitters, using a technically complete Fowler-Nordheim-type equation**  [Version scientifically equivalent to corrected proof]

**Richard G. Forbes**

Advanced Technology Institute, Faculty of Engineering and Physical Sciences,

University of Surrey, Guildford, Surrey GU2 7XH, UK

E-mail: r.forbes@trinity.cantab.net

**Abstract**

In papers on cold field electron emission from large area field emitters (LAFEs), it has become widespread practice to publish a misleading Fowler-Nordheim-type (FN-type) equation. This equation over-predicts the LAFE-average current density by a large highly-variable factor thought to usually lie between $10^3$ and $10^9$. This equation, although often referenced to FN's 1928 paper, is a simplified equation used in undergraduate teaching, does not apply unmodified to LAFEs, and does not appear in the 1928 paper. Technological LAFE papers often do not cite any theoretical work more recent than 1928, and often do not comment on the discrepancy between theory and experiment. This usage has occurred widely, in several high-profile American and UK applied-science journals (including *Nanotechnology*), and in various other places. It does not inhibit practical LAFE development, but can give a misleading impression of potential LAFE performance to non-experts. This paper shows how the misleading equation can be replaced by a conceptually complete FN-type equation that uses three high-level correction factors. One of these, or a combination of two of them, may be useful as an additional measure of LAFE quality; this paper describes a method for estimating factor values using experimental data, and discusses when it can be used. Suggestions are made for improved engineering practice in reporting LAFE results. Some of these should help to prevent situations arising whereby an equation appearing in high-profile applied-science journals is used to support statements that an engineering regulatory body might deem to involve professional negligence.

## 1. Introduction

### 1.1 General background

In a recent paper [1] in *Nanotechnology*, Arif et al. presented interesting results about cold field electron emission (CFE) from arrays of metal post-like emitters grown on a flexible graphene-like substrate. This is exciting technological progress with various potential applications.

Their article uses an equation equivalent to eq. (13) below, which is related to an equation developed by Fowler and Nordheim (FN) in 1928 [2-6]. Equation (13) appears widely in modern CFE technological literature. It is often used, as in Ref. [1], to derive a formula [eq. (16) below] that is used to estimate the characteristic field enhancement factor for a large-area field emitter (LAFE).

Equation (16) is adequate as a basic approximation. However, eq. (13) can be seriously misleading when applied to emitter arrays or other forms of LAFE, since it over-predicts [7] the "macroscopic" (i.e., LAFE-average) emission current density (ECD), by a factor thought to typically lie between $10^3$ and $10^9$. This occurs because the conceptually complete FN-type equation for CFE from relevant LAFEs contains three important correction factors not present in eq. (13).

This equation misuse is not dangerous when experimental results are clearly reported. However, there is a real possibility that non-experts (particularly those motivated to fund nanotechnology development) may take this widely published misleading equation out of context, and reach spurious conclusions about potential LAFE performance. The present paper aims to replace eq. (13) by conceptually more complete equations for LAFE-average ECD ($J_M$), and to show that this brings advantages.

In technological LAFE papers, often the main or only theoretical reference is to the FN 1928 treatment of local ECD ($J_L$). New material below defines and discusses a macroscopic pre-exponential correction factor $\lambda_M$, and suggests that it could be called the "LAFE performance factor". However, it has seemed clearest to develop this new approach as part of an account of all post-1928 theoretical developments directly relevant to LAFE data interpretation.

The paper's structure is as follows. Section 2 gives basic definitions, and outlines how FN-type equations have become increasingly complete as local descriptions of CFE from metals. Section 3 shows how FN-type equations are modified to apply to LAFEs. Section 4 considers how to extract LAFE characterisation parameters from experimental current-voltage measurements. Section 5 illustrates use of the resulting formulae. Section 6 provides discussion and recommendations for improved practice. An appendix contains mathematical details. A list of acronyms used is provided as electronic supplementary material. Technical conventions follow Refs. [4, 5]; in particular, values of universal constants are given to seven significant figures.

## 2. Technical and historical background – local current densities

### 2.1 Basics

Field-assisted electron tunnelling through an exact or rounded triangular barrier is often known as *Fowler-Nordheim (FN) tunnelling*. *Cold field electron emission* (CFE) is the emission regime where: (a) the electrons in the emitting region are in thermodynamic equilibrium (or very nearly so); and (b) most emitted electrons escape by FN tunnelling from states close to the Fermi level and well below the barrier maximum. Many practical field-assisted emission devices, including the Ref. [1] arrays, operate in the CFE regime.

CFE from bulk metals is described by a family of approximate equations called *Fowler-Nordheim-type (FN-type) equations*. These equations were derived for material situations where emission comes from a degenerate metal-like conduction band of sufficient depth, and quantum-confinement effects [8,9] do not operate. The term *bulk*, used above, implies that quantum-confinement effects need not be considered.

These limitations mean that, strictly, physical FN-type equations are valid only for bulk metals, and (in some circumstances) for bulk crystalline semiconductor conduction bands that are degenerate near the emitting surface, as a result of field penetration [10,11]. Section 4.4 discusses their application in other situations.

Different FN-type equations may derive from different physical emission assumptions or from different mathematical approximations. Different equations apply to an emitter surface location and to an average quantity relating to a LAFE. Consequently, different FN-type equations may yield "current densities" differing by factors of up to $10^9$ or more.

There are many (perhaps 20 or more) different FN-type equation variants in the literature, some of which are confusingly referred to (in some articles in which they occur) as *the* Fowler-Nordheim equation or model. Double or multiple conflicting meanings exist for the basic terms "field", "field enhancement factor" and "current density", and double or multiple conflicting meanings exist for some letter-symbols. The use of a misleading FN-type equation needs to be seen against the background of this wider terminological disarray.

## 2.2 The original and elementary FN-type equations

The *original* FN-type equation was developed in 1928 by Fowler and Nordheim [2,3], using several hypotheses and some simplifying assumptions. Thus FN: (1) assumed a smooth, flat, planar emitter surface, with constant electric field outside it; (2) disregarded details of emitter atomic structure; (3) ignored what would now be called the exchange-and-correlation interaction between the escaping electron and the emitter surface [12,13], and assumed tunnelling took place through an exact triangular (ET) barrier; (4) assumed a Sommerfeld-type free-electron model [14,15] for the emitter electron states (and improved it by including electron spin); (5) assumed electrons obeyed Fermi-Dirac statistics; (6) approximated emitter temperature as 0 K; and (7) made various mathematical approximations. Their resulting equation gives the *local emission current density* (LECD) $J_L$ in terms of the local thermodynamic work-function $\phi$ and the local surface electric field $F_L$ (called here the *local barrier field*). In modern notation, their eq. (21) is conveniently written

$$J_L = P_F^{FN} a \phi^{-1} F_L^2 \exp[-b\phi^{3/2}/F_L] \qquad (original, for\ LECD), \qquad (1)$$

where $a$ [$\cong$ 1.541 434 µA eV V$^{-2}$] and $b$ [$\cong$ 6.830 890 eV$^{-3/2}$ V nm$^{-1}$] are universal constants sometimes called the First and Second FN Constants [5], and $P_F^{FN}$ is a tunnelling pre-factor (see Refs. [2, 4, 16]). The term *local* means "applicable to a particular *lateral* location on the emitter surface".

Teaching about FN tunnelling does not use FN's full quantum-mechanical approach, because the simple-JWKB (Jeffreys-Wentzel-Kramers-Brillouin) approximate method [16, 17] is easier to understand. When applied to FN's exact triangular barrier, the simple-JWKB method generates an approximate formula [17, 18] without the tunnelling pre-factor. This *elementary* FN-type equation for LECD (for $J_L$ in terms of $\phi$ and $F_L$) is

$$J_L = a\phi^{-1} F_L^2 \exp[-b\phi^{3/2}/F_L] \qquad (elementary, for\ LECD). \qquad (2)$$

## 2.3 The zero-temperature Murphy-Good FN-type equation

An electron leaving a surface experiences forces due to both (a) the electric field that would exist in its absence, and (b) the electric field due to the surface reaction to the electron presence outside it. This reaction, now formally known as an *exchange-and correlation (E&C) effect* [12, 13], is the quantum-mechanical generalization of classical image-force effects.

FN knew that exact triangular barriers were physically unrealistic, and would be rounded by Schottky's [19] planar image effect, to give barriers now called Schottky-Nordheim (SN) barriers [19-23]. However, they seriously underestimated the amount of rounding (see Fig. 1). Thus, their remarks on how rounding affects transmission probability $D$ are incorrect and misleading. In late 1928, Nordheim [20] tried to calculate $D$ for the SN barrier. Unfortunately, his elliptic-function mathematics was incorrect, and significantly under-predicted the increase in $D$.

FIGURE 1 NEAR HERE

Appropriate corrections were introduced into mainstream CFE theory in the 1950s [24-27], most definitively by Murphy and Good (MG) [27]. The outcome was the revised FN-type equation

$$J_L = t_F^{-2} a\phi^{-1} F_L^2 \exp[-v_F b\phi^{3/2}/F_L] \quad \textit{(standard or Murphy-Good, for LECD)}, \quad (3)$$

where $v_F$ ("vee$_F$") and $t_F$ are particular values of well-defined mathematical functions $v$ and $t$ [22, 27]. Mathematically, $v$ is now known to be a special solution [28] of the Gauss hypergeometric equation, and $t$ is obtained from $v$; $t$ and $v$ are sometimes called "field emission elliptic functions", but are better called *SN barrier functions* [22], because they apply only to tunnelling through a SN barrier.

Equation (3) predicts LECD ($J_L$) values 100 or more times greater than those given by eqns (1) or (2). For example, for the typical metal work-function value $\phi$ = 4.5 eV, and for $F_L$ = 4 V/nm, eq. (2) predicts $J_L \approx 2.4 \times 10^5$ A/m$^2$, but eq. (3) predicts $J_L \approx 1.2 \times 10^8$ A/m$^2$, a difference by around 500.

From the 1960s, this *standard* (or *zero-temperature Murphy-Good*) FN-type equation for LECD was widely used to interpret CFE experimental data, in particular from Spindt arrays [29]. However, apart from the barrier-form difference, most of FN's assumptions were also used by MG. Further, MG used a JWKB-like approximation [26, 27] that does not generate the tunnelling pre-factor that should physically be present [16, 18, 30], although a pre-exponential correction factor ($t_F^{-2}$) relating to analytical integration over emitter electron states does appear.

## 2.4 More about exchange-and-correlation effects

Because the 1950s work uses Schottky's classical planar image potential energy (PE) to model E&C effects, but Schottky's theory was derived (in 1914) for a good classical electrical conductor, the 1950s corrections apply *in detail* only to bulk metals and (slightly modified) to bulk semiconductors. In the 1990s, technological work on LAFEs dealt mainly with carbon-based materials [31, 32]; however, the E&C interactions between an escaping electron and carbon-based emitters are not the same as for metals, and were not clearly understood in the 1990s. (It now seems [33] that for carbon nanotubes they may be weaker than the classical image PE.) In the absence of detailed knowledge, there was some justification for using the elementary rather than the Murphy-Good equation.

However, much recent LAFE work concerns semiconducting nanowires, and Ref. [1] deals with metal nanowires. For these materials, the E&C interactions must be included.

There is a question of whether the classical image PE is a good model for E&C interactions with metals and semiconductors. The classical model is inappropriate very close to the surface (not least because it goes down to an unphysical minus infinity), but the SN barrier peak is often around 1 nm from the emitter surface. At this distance, the classical image PE should be adequate, certainly in JWKB-type treatments; thus, for metals, and as an approximation for semiconductors, the SN-barrier model should be better than the ET-barrier model. There is theoretical evidence [13, 34] and some experimental evidence [35] to support this conclusion.

The 1950s modifications, to correct the treatment of the SN barrier in FN-type equations, are not as well known as they might be. MG's paper [27] is deeply mathematical and difficult to follow. Ref. [22], supported by Refs. [28, 36-41] provides a modern reformulation of SN barrier theory that aims to be easier to understand, but contains some typographical errors. In addition to those already corrected [23], eq. (5.1) in Ref. [22] should read: $u(l') = -dv/dl' \approx 1 - q - q \ln l'$.

## 2.5 The technically complete FN-type equation for local emission current density

Modern emitters often have tip radii smaller than those of 1960s emitters. For sufficiently small tip radii the SN barrier ceases to be a good model, due to field-fall-off effects, and the correction factor $v_F$ must be replaced by a more general "barrier form correction factor" $\nu_F$ ("nu$_F$"). For any well-behaved barrier model, values of $\nu_F$ can be calculated numerically, using JWKB-type methods.

The effect of abandoning other simplifying assumptions and approximations in the FN 1928 paper can be represented *conceptually* by replacing $t_F^{-2}$ in eq. (3) by a much more general *local pre-exponential correction factor*, denoted here by $\lambda_L$. The resulting equation, called here the *technically complete* FN-type equation for LECD (for $J_L$ in terms of $\phi$ and $F_L$), is

$$J_L = \lambda_L a\phi^{-1} F_L^2 \exp[-v_F b\phi^{3/2}/F_L]. \qquad \textit{(technically complete, for LECD)}. \qquad (4)$$

When examined in detail [42], the correction factor $\lambda_L$ decomposes into the product of an electron supply correction factor $\lambda_Z$ and a tunnelling pre-factor $P_F$, and $\lambda_Z$ itself further decomposes into the product of a correction factor $\lambda_{D0}$ that relates to the process of summing over electron states in a zero-temperature free-electron model, a correction factor $\lambda_T$ that relates to the effects of temperature in a free-electron model (in fact discussed by Murphy and Good [27]), and a correction factor $\lambda_E$ that relates to use of atomic-level wave-functions and realistic electron band-structure (e.g., Refs. [10, 43]). In turn, the correction factor $\lambda_{D0}$ breaks down into the product of several factors [41], one of which is the factor $t_F^{-2}$ in eq. (3). Details of these decompositions are not important in technological contexts, but this outline can provide understanding of how uncertainty in $\lambda_L$ is estimated.

$P_F$, $\lambda_{D0}$ and $\lambda_T$ can be estimated satisfactorily for free-electron models of planar emitters and are of order unity. This can be assumed to hold for curved emitters, too, provided tip radius is "not too small". Most of the uncertainty in $\lambda_L$ arises from the intense difficulty of applying accurate quantum mechanics to real field emitters, even using simplified surface models. My present best guess [42], which incorporates the views of Modinos [43, 44], is that $\lambda_L$ most likely lies between 0.005 and 10, and thus $\lambda_L^{-1}$ between 0.1 and 200. This guess is built from the range of values likely for the correction factors $P_F$, $\lambda_{D0}$ and $\lambda_T$, and the likely uncertainty over $\lambda_E$. Table 1 shows relevant information.

TABLE 1 NEAR HERE

Specific approximate versions of eq. (4), other than the familiar ones recorded as eqns (1) to (3), may sometimes be useful. Thus, the *full FN-type equation for the SN barrier, for LECD $J_L$*, is obtained by keeping $\lambda_L$ but replacing $v_F$ by $\bar{v}_F$.

Equation (4) (and all physical FN-type equations as defined here) are strictly applicable only if the emitter tip radius is "not too small". My present expectation is that quantum-confinement effects may influence field emitted energy distributions if the tip dimension normal to the emitting surface is less than about 20 nm [9], but that local current densities will be significantly affected only for tip dimensions around 1 to 2 nm and below [8]. For tip sizes near and below this, physical FN-type equations may need replacing by one or more new families of CFE equations in which (a) $\phi$ is replaced by a more general reference barrier height $H_R$ [8], (b) different forms may be needed for the pre-exponential (and for the exponent, if the tip radius is less than 1 nm [8]), and (c) a series of terms, rather than a single term, may be needed. Broadly similar changes would be needed in CFE equations for non-metals, if any tip dimension is small (as with many carbon nanotubes).

### 3. Fowler-Nordheim-type theory for large area field emitters (LAFEs)

#### 3.1 Macroscopic field enhancement factors

Older CFE work, with single-point-geometry emitters, used auxiliary equations to relate a characteristic barrier field ($F_C$) to the applied voltage $V$, and the corresponding characteristic local emission current density ($J_C$) to the total emission current $i$. LAFE papers normally use different variables, namely the *macroscopic field* denoted here by $F_M$ and the *macroscopic* or *LAFE-average current density* denoted here by $J_M$. Auxiliary equations are needed that relate these to values of $F_C$ and $J_C$ characteristic of the LAFE as a whole. There is also a separate issue of how to correctly "pre-convert" measured voltages and currents into values of $F_M$ and $J_M$; this is straightforward when there is no current saturation or voltage drop due to series resistance in the measuring circuit.

LAFE geometries can be decomposed into structure on two scales. The first is the *base structure*, often a planar substrate with a distant planar counter-electrode, but sometimes a curved substrate—as in Ref. [1]—with a distant counter-electrode of unspecified shape. The second-scale structure is some form of "pointy" local structure—in Ref. [1] the array of posts.

The *macroscopic field* $F_M$ is defined here as the field that would exist on the base structure *in the absence of* the second-scale structure. (A different parameter, the *gap field* $F_G$, is defined as the mean field between the counter-electrode and emitting region at the top of the second-scale structure.)

When the base and counter-electrode are parallel planes, separated by distance $Z_p$, with a voltage of magnitude $V$ applied between them, $F_M$ is given by

$$F_M = V/Z_p. \tag{5}$$

This formula is sometimes used to *define* $F_M$, but the more careful definition above seems better.

As is well known, a pointed structure enhances the macroscopic field. For a LAFE, the relevant characteristic barrier field $F_C$ is formally related to $F_M$ by a *characteristic macroscopic field enhancement factor* (characteristic MFEF), denoted here by $\gamma_C$ and defined by:

$$\gamma_C \equiv F_C / F_M. \tag{6}$$

(MFEFs are often denoted by $\beta$ in LAFE literature, but $\beta$ has a different meaning in many older CFE papers, thus $\gamma$ is preferred here.)

The value of $\gamma_C$ depends on the electrostatics of the whole system. Thus, $\gamma_C$ is affected by the *gap-length* between the emitter and counter-electrode. Most LAFEs operate with gap-length much greater than the maximum height $h_{max}$ of the second-scale structure. In this case, $\gamma_C$ for a given LAFE has a well-defined *large-gap value* sometimes denoted by $\gamma_\infty$. It is usually $\gamma_\infty$ that is of technological interest. Due to *mutual screening* effects [45], the characteristic MFEF for an array of identical "features" may be smaller than that for an individual feature considered in isolation.

An alternative approach [46, 47] uses the gap field and a "gap field enhancement factor" (GFEF). However, for large gap-lengths, the characteristic GFEF tends to the same limiting value $\gamma_\infty$ as the characteristic MFEF for the geometry in question. This paper prefers to use MFEFs.

**3.2 Area efficiency of emission (AEE)**

A second need, for LAFEs, is to relate the LAFE-average current density $J_M$ to the characteristic LECD $J_C$ associated with the characteristic barrier field $F_C$. Unnecessary problems are avoided by considering parallel-plane geometry. Figure 1 in Ref. [1] shows that the total emitting area at the tips of the cylindrical posts must be significantly less than the total area (*macroscopic area, or* "footprint") $A_M$ of the base on which they stand. Thus, $J_M$ must be much less than the LECDs at the post tips. This suggests the auxiliary equation

$$J_M = \alpha_M J_C . \tag{7}$$

$\alpha_M$ is the *area efficiency of emission (AEE)*, and is very much less than unity.

A formal definition [7] is as follows. For a LAFE, the LECD $J_L$ varies strongly with location on the surface of the complete emitter, but will be highest for the strongly emitting regions at the tips of pointed structures. The total emission current $i$ is obtained by integrating $J_L$ over the whole emitter surface area; thus $i = \int J_L \mathrm{d}A$. Now consider the surface location at which $J_L$ has its maximum across-surface value (for any given applied voltage), and designate this maximum value as the LAFE characteristic value $J_C$. A *notional emission area* $A_n$ can then be defined by

$$A_n \equiv [\int J_L \mathrm{d}A] / J_C . \tag{8}$$

In principle, other methods of defining $F_C$ and $J_C$ could be used. The procedure above is a slight generalization, for LAFEs, of that used [48] to define notional emission area for single-point-geometry emitters; this, in turn, is a formal version of the procedure used by Stern, Gossling and Fowler in 1929 to define what they called a "weighted mean area" (see footnote on p. 700 of Ref. [3]).

For a parallel-plane LAFE, the area efficiency of emission (AEE) $\alpha_M$ is then given by

$$\alpha_M = A_n / A_M = J_M / J_C . \tag{9}$$

Thus, $\alpha_M$ can also be interpreted as the current-density ratio $J_M/J_C$.

### 3.3 The technically complete FN-type equation for LAFEs

When both auxiliary equations are combined with eq. (4), the outcome is the *technically complete FN-type equation for LAFEs* (for $J_M$ in terms of $\phi$ and $F_M$):

$$J_M = \alpha_M \lambda_C a\phi^{-1} \gamma_C^2 F_M^2 \exp[-v_F b\phi^{3/2}/\gamma_C F_M] \qquad (\textit{technically complete, for LAFEs}), \qquad (10)$$

where $\lambda_C$ is the characteristic value of $\lambda_L$. Usually, $\gamma_C$ would be $\gamma_\infty$. For CFE from LAFEs with metal emitting regions, if no quantum-confinement occurs, eq. (10) is sufficiently general that it *conceptually* covers all other known physical effects that influence $J_M$.

It is difficult to establish accurate values individually for $\alpha_M$ and $\lambda_C$; hence, it can be better to combine them into a *macroscopic pre-exponential correction factor (MPCF)* $\lambda_M$ given by

$$\lambda_M \equiv \alpha_M \lambda_C = J_M / \lambda_C^{-1} J_C. \qquad (11)$$

With a curved substrate, the detector area that receives a measured current falls is larger than for parallel-plane geometry, and definitions may need slight modifications. For simplicity, plane-parallel geometry is assumed here.

If reliable values of the MPCF ($\lambda_M$) or AEE ($\alpha_M$) could be established experimentally, then one or both could be useful as engineering parameters-of-merit for describing array behaviour. For example, one could envisage engineering trade-offs in LAFE performance between high $\gamma_\infty$ (low onset voltage) and high $\lambda_M$ (high LAFE-average current density) arrays.

If the barrier is taken as a SN barrier, we get the *full SN barrier equation for LAFEs*

$$J_M \approx \lambda_M a\phi^{-1}\gamma_C^2 F_M^2 \exp[-v_F b\phi^{3/2}/\gamma_C F_M] \qquad \textit{(full SN barrier equation for LAFEs)}, \qquad (12)$$

In practice, this is the most useful equation for analysis of LAFE behaviour.

In eq. (10), if the correction factors $v_F$, $\lambda_C$ and $\alpha_M$ are all set equal to unity, this gives the *elementary FN-type equation for LAFEs* (for $J_M$ in terms of $\phi$ and $F_M$):

$$J_M = a\phi^{-1}\gamma_C^2 F_M^2 \exp[-b\phi^{3/2}/\gamma_C F_M] \qquad \textit{(elementary, for LAFEs)}. \qquad (13)$$

This equation is equivalent to eq. (1) in Ref. [1], and is the one commonly given in LAFE literature. As compared with eq. (13), the conceptually complete eq. (10) contains three correction factors $v_F$, $\lambda_C$ and $\alpha_M$. The most serious theoretical defect of many LAFE papers is the omission of $\alpha_M$, as can be shown by estimating its likely value.

### 3.4 Preliminary estimation of area efficiency of emission (AEE)

Finding AEE values directly is difficult, because simultaneous measurement of $J_M$ and $J_C$ is difficult, but one can compare typical values. For a tungsten single-point-geometry emitter, Dyke and Trolan [49] reported a working $J_C$-range lying between about $10^5$ A/m$^2$ and about $6\times10^{10}$ A/m$^2$. Such values are typical of metals. Specific $J_M$-values reported for the gold nanopost array of Ref. [1] lie between 0.1 A/m$^2$ and 3 A/m$^2$. Such values are typical of those obtained from many LAFEs [50], and should correspond to some $J_C$-value in the Dyke-Trolan range. Thus, one can infer (for metals) that $\alpha_M$ can sometimes be smaller than (3A/m$^2$)/(10$^5$ A/m$^2$), i.e. $3\times10^{-5}$. This establishes unambiguously that the area efficiency of emission needs to appear in CFE theory. The same conclusion is reached using the geometrical arguments earlier, or electrostatic arguments about mutual screening [7].

## 4. Extraction of characterization parameters from orthodox emission data

A field emission situation (and related data) can be described as *orthodox* when certain and physical and mathematical requirements are satisfied. The former are that (a) emission is controlled solely by the tunnelling barrier at the emitter/vacuum interface, and (b) the physical macroscopic field enhancement factor $\gamma_C$ is physically independent of the measured voltage. The latter are that (c) the emitter/vacuum tunnelling barrier can be adequately modelled as an SN barrier of constant zero-field height (normally equal to the assumed emitter work function), and (d) the so-called "conventional slope-analysis assumption" discussed below is adequately satisfied. The physical requirements imply, amongst other things, that there are no internal-voltage-drop, space-charge or field-dependent-geometry effects.

This section describes a method for extracting characterisation parameters from *orthodox* experimental data; related mathematics is in Appendix A. In reality, it is probable that many technologically interesting emission situations are not orthodox. Appendix A also sets out a method for testing whether an experimental FN plot describes orthodox data. If the emission is not orthodox, then the extraction procedure described here needs modifying, and the interpretation of extracted parameters may involve complicated issues; unorthodox emission situations are beyond the scope of the present paper and will be discussed elsewhere, later.

### 4.1 Improved analysis of FN-plot slope

The slope ($S_M$) of a FN plot of type [$\ln\{J_M/F_M^2\}$ vs $1/F_M$] is defined by

$$S_M \equiv \partial \ln\{J_M/F_M^2\} / \partial(1/F_M) . \tag{14}$$

Applying definition (14) to eq. (13) gives (if $\phi$ and $\gamma_C$ are constant)

$$\tag{15}$$

$$S_\text{M} = -b\phi^{3/2}/\gamma_\text{C}.$$

A slope characterisation parameter, denoted here by $\gamma_\text{C}^{*1}$, can be extracted via eq. (15) as

$$\gamma_\text{C}^{*1} = -b\phi^{3/2}/S_\text{M}^\text{expt}, \qquad (16)$$

where $S_\text{M}^\text{expt}$ is the slope of the regression line fitted to the experimental FN plot.

If definition (14) is applied, instead, to eq. (10) or (12), then the result can be written

$$S_\text{M} = -\sigma_\text{M} b\phi^{3/2}/\gamma_\text{C} \qquad (17)$$

where $\sigma_\text{M}$ is a *generalized slope correction factor,* possibly dependent on many variables. The corresponding characterisation parameter is denoted by $\gamma_\text{C}^{*\sigma}$ and given by

$$\gamma_\text{C}^{*\sigma} = -\sigma_\text{M} b\phi^{3/2}/S_\text{M}^\text{expt} = \sigma_\text{M} \gamma_\text{C}^{*1}. \qquad (18)$$

When emission is orthodox, $\gamma_\text{C}^{*\sigma}$ can be interpreted as the value of a conventional (voltage-independent) characteristic field enhancement factor (MFEF), and $\gamma_\text{C}^{*1}$ as an approximation to $\gamma_\text{C}^{*\sigma}$.

Most empirical MFEF values in the literature have been deduced by setting $\sigma_\text{M}=1$. All such values (for example, those in Ref. [50]), are technically in error, even for orthodox emission. The issue is: how badly? The problem is that definition (14) will create for $\sigma_\text{M}$ an expression involving a summed series of terms, one each for any dependence on $F_\text{M}$ in each of the parameters $\alpha_\text{M}$, $\lambda_\text{E}$, $\lambda_T$, $\lambda_{D0}$, $P_\text{F}$, $\phi$, $\gamma_\text{C}$ and $\nu_\text{F}$, for each time the parameter appears in eq. (10). Some terms will be negligible or small, but the situation is less obvious for others (in particular, for $\alpha_\text{M}$, $\phi$, $\lambda_\text{E}$, $\gamma_\text{C}$ and $\nu_\text{F}$). So far, there has been little systematic investigation. The *conventional slope-analysis assumption* is that, when calculating $\sigma_\text{M}$, only the direct dependence on $1/F_\mathrm{M}$ and the dependence of $v_\mathrm{F}$ on $F_\mathrm{M}$ need to be taken into account.

The best established result relates to eq. (12) and applies in the case of assumed orthodox emission (where an SN barrier is assumed). If one neglects $F_\mathrm{M}$-dependence in all parameters other than $v_\mathrm{F}$, then $\sigma_\mathrm{M}$ is given by the mathematical SN barrier function $s$ [22, 24]. This typically has a mid-operating-range value around 0.955, which would imply that $\gamma_\mathrm{C}^{*1}$ over-predicts $\gamma_\mathrm{C}^{*\sigma}$ by around 4.5%. Obviously, if error is less than around 10%, then this has limited technological significance, but perhaps uncorrected MFEF values (particularly $\gamma_\mathrm{C}^{*1}$ values) should usually not be stated to a precision better than two significant figures.

Preliminary indications exist [51] that errors might be significantly worse for LAFEs where the most strongly emitting features have tips with radius of curvature below about 20 nm. More generally, we do not know how valid the conventional slope analysis assumption actually is.

### 4.2 Estimation of macroscopic pre-exponential correction factor $\lambda_\mathrm{M}$

Estimates of the MPCF $\lambda_\mathrm{M}$ can be made using eq. (11): $J_\mathrm{M}$ can be obtained experimentally, and $\lambda_\mathrm{C}^{-1}J_\mathrm{C}$ estimated with the help of theory. In the case of orthodox emission, an FN plot is used to obtain an estimate of the field enhancement factor $\gamma_\mathrm{C}$. One then chooses a suitable linked pair ($F_\mathrm{Mw}$ and $J_\mathrm{Mw}$) of experimental values of $F_\mathrm{M}$ and $J_\mathrm{M}$: $F_\mathrm{Mw}$ should be chosen so that $1/F_\mathrm{Mw}$ is near the middle of the range of $(1/F_\mathrm{M})$-values used when finding $\gamma_\mathrm{C}$. An SN barrier is assumed. ($\lambda_\mathrm{C}^{-1}J_\mathrm{C}$) can then be estimated as described in Appendix A, and $\lambda_\mathrm{M}$ found from eq. (11) by putting $J_\mathrm{M}= J_\mathrm{Mw}$. Any uncertainty over the true value of $\sigma_\mathrm{M}$ will lead to uncertainty in this estimate of $\lambda_\mathrm{M}$.

### 4.3 Estimation of area efficiency of emission $\alpha_\mathrm{M}$

The macroscopic pre-exponential correction factor (MPCF) $\lambda_M$ is a "provisional estimate" of the area efficiency of emission (AEE) $\alpha_M$. In principle, a best guess at the range of values within which $\alpha_M$ actually lies is obtained by inverting eq. (11) to give

$$\alpha_M = \lambda_C^{-1} \lambda_M, \tag{19}$$

and then combining any uncertainty in $\lambda_M$ with the uncertainty in $\lambda_C^{-1}$ (which is the inverse of uncertainty in $\lambda_L$ noted in Table 1). This procedure is illustrated in Section 5.

If an AEE value $\alpha_M$ can be estimated, and the LAFE macroscopic area $A_M$ is known, then the notional emission area is $A_n = \alpha_M A_M$, with uncertainty factors the same as for $\alpha_M$.

In principle, an alternative method for estimating AEE values would be to derive $A_n$ from the intercept of an *i-V*-type FN plot, and take $\alpha_M = A_n/A_M$. Uncertainties similar to those just discussed would arise, but in a different way. In practice, existing theory for interpreting FN-plot intercepts is not adequate, and needs revising [51].

**4.4 Formulae for emitters that do not obey physical Fowler-Nordheim-type equations**

The theory above has been derived using FN-type equations; strictly, these apply *physically* to "bulk" metal emitter arrays (i.e., when the tip of each emitter is "not too small"). Ideally, the theory needs generalization to cover other LAFE emission situations. However, this is not straightforward and unlikely to happen quickly. This is because theory for metal emitters is itself far from complete, and developing this further seems the best path forwards (in particular, the disregarded terms in $\sigma_M$ merit exploration).

The main difference between bulk metals and other orthodox emitting materials lies in the electronic-band-structure effects that go into the electronic-structure correction factor $\lambda_E$, and hence into $\lambda_L$ and $\lambda_M$. The barrier-form effects that go into the correction factor $\nu_F$ are broadly similar for different materials (though not the same in detail), and for all materials there is a need to incorporate the AEE $\alpha_M$ into an equation for LAFE-average current density $J_M$.

The most serious deficiency of existing LAFE theory is the omission of $\alpha_M$ from equations for $J_M$. As a temporary expedient, it seems acceptable engineering practice to use eq. (12) to describe emission from orthodox emitters that are not bulk metals. More generally, for non-orthodox emitters one can argue that eq. (12) can be used as an empirical fitting equation—certainly as a temporary expedient. (But physical interpretations then need to be established for the fitting parameters "$\gamma_C$" and "$\lambda_M$"). In all cases, using eq. (12) would certainly be better than the present practice of using an elementary FN-type equation.

### 4.5 The need for well-defined engineering parameters-of-merit

The formulae above represent best practice in the present state of confirmed knowledge. As knowledge improves, details (in particular, $\sigma_M$-values) will change. Thus, there will be changes in values of $\gamma_C^{*\sigma}$, as derived (for given data) using eq. (18), and possibly in best estimates of MPCF ($\lambda_M$) and AEE ($\alpha_M$) values.

For engineering comparisons between LAFEs fabricated from different materials, it is probably better at present to compare estimates derived in a standard way, rather than estimates derived in different ways as knowledge improves. I thus suggest that technological papers should continue to quote the extracted value of $\gamma_C^{*1}$, but should indicate that values are derived from the general formula (17) by using $\sigma_M = 1$.

It also seems important to establish whether or not the measured emission is orthodox (for instance, by the test in Appendix A), and to report the result, as this may affect the physical interpretation of $\gamma_C^{*1}$.

Similar consistency considerations apply to $\lambda_M$ but generate a slightly different proposal. For orthodox emission, limited self-consistency is achieved by using Appendix A to find a value for $\lambda_C^{-1} J_C$, and hence a value for $\lambda_M$ via eq. (11). This approach takes $\sigma_M$ equal to $s_{Cw}$ as given by eq. (A12), which differs from unity. I suggest that appropriate engineering practice would use the Appendix A approach, and report the value of $s_{Cw}$ involved.

Obviously, this means that different values of $\sigma_M$ are being used to generate estimates of $\gamma_C$ and $\lambda_M$. This is not entirely self-consistent, but seems the most practical proposal at present.

## 5. Numerical illustration of extraction procedure for orthodox emission

Appendix A describes the extraction procedure. Published papers (including Ref. [1]) often do not contain all the experimental information necessary to apply it fully. Simulated results, shown as a $J_M$-$F_M$-type FN plot in Fig. 2, are used here to illustrate and validate it. The simulation assumes orthodox emission (including an SN-barrier), and uses the input-parameter values $\phi = 4.5$ eV, $\gamma_C = 500$, $\lambda_C = 1$ and $\alpha_M = 10^{-9}$ (hence $\lambda_M = 10^{-9}$). Data points have been calculated for values of characteristic scaled barrier field $f_C$ lying at intervals of 0.005 in the range 0.25 to 0.34 (see Appendix A for the definition of $f_C$). The FN plot is analyzed as below. Results are given to 3 significant figures for working purposes, but are not physically accurate to this precision.

FIGURE 2 NEAR HERE

The slope $S_M^{\text{expt}}$ derived from ruler-based measurements on Fig. 2 is 123 Np V μm$^{-1}$, where the neper (Np) is used as the unit of natural-logarithm difference. The working point "w" is chosen to correspond to the logarithm value −34. From Fig. 2, the related value $1/F_{Mw}$ is found as 0.126 μm/V, the related field as $F_{Mw} \approx 7.96$ V/μm, and the related macroscopic ECD as $J_{Mw} \approx 0.109$ A/m$^2$.

From eqns (A9), (A11), (A12), we obtain $f_{Cw}^{*1} \approx 0.301$, $f_{Cw} \approx 0.286$, $s_{Cw} \approx 0.952$. The scaled barrier field value $f_{Cw} \approx 0.286$ is within the acceptable range of values (0.22 to 0.32) noted in Appendix A.

From eq. (16) we extract $\gamma_C^{*1} \approx 531$; from eq. (18), with $\sigma_M = s_{Cw} = 0.952$, we get $\gamma_C^{*\sigma} \approx \gamma_C^{*0.952} \approx 506$. This compares well with the input value 500 (exact agreement would be expected only if, by luck, the chosen working point corresponds exactly to the value of $1/F_M$ at which, in a FN plot, the tangent to the theoretical curve is parallel to the fitted regression line [38]). This result is one demonstration that the procedure is self-consistent, and also shows that $\gamma_C^{*\sigma}$ is a better parameter than $\gamma_C^{*1}$.

From eqns (A3) to (A5), with $\phi = 4.5$ eV, and with $\eta^{SN}$ and $\theta^{SN}$ evaluated as 4.64 and $6.77 \times 10^{13}$ A/m$^2$, respectively, and using $f = f_{Cw} = 0.286$, we obtain the working-point value of $\lambda_C^{-1} J_C$ as $1.41 \times 10^8$ A/m$^2$. From eq. (11), with $J_{Mw} = 0.109$ A/m$^2$, as deduced above, we extract $\lambda_M \approx 7.7 \times 10^{-10}$. This compares adequately with the input value of $10^{-9}$ (exact agreement is not expected), and is a second demonstration of procedure self-consistency.

The simulation took $\lambda_C = 1$, so eq. (11) yields the value $\alpha_M \approx 7.7 \times 10^{-10}$. In real situations, this result needs to be taken as a "provisional estimate" of $\alpha_M$. The uncertainties indicated in Table 1 would apply, and (for $\lambda_M \approx 7.7 \times 10^{-10}$) would predict that $\alpha_M$ lies in the range $7.7 \times 10^{-11}$ to $1.5 \times 10^{-7}$.

In principle, uncertainty over the prediction of $\lambda_M$ implies that the uncertainty range for $\alpha_M$ needs to be expanded. Lack of necessary theory means the size of this additional uncertainty cannot be reliably assessed, at present.

## 6. Discussion

### 6.1 Summary of problems with using the elementary FN-type equation

The main problems with using the elementary equation (13) to describe the LAFE-average current density $J_M$ can be summarised as follows.

(1) Equation (13) does not include the barrier form correction factor $\nu_F$. This omission, by itself, causes eq. (13) to under-predict $J_M$ by a significant factor, of order 100 for a SN barrier.

(2) Equation (13), understandably, does not include the characteristic local pre-exponential correction factor $\lambda_C$, or any of the factors that it decomposes into. This omission has some effect on current-density prediction; inclusion of $\lambda_C$ makes a more complete theory and facilitates discussion of uncertainties.

(3) Equation (13) does not include the area efficiency of emission $\alpha_M$. This is a serious omission and causes eq. (13) to over-predict $J_M$ by a very large factor, perhaps as much as $10^9$ or more.

(4) Equation (15) for FN-plot slope $S_M$ does not include the correction factor $\sigma_M$. For orthodox emission this correction is expected to be small (less than 10%), and not easily detected in experiments.

These problems exist whether emission is orthodox or not. Additional problems, beyond the scope of this paper, exist when emission is not orthodox.

## 6.2 Origins of existing practice

The omission of the area efficiency of emission $\alpha_M$ in the FN-type equations applied to LAFEs is widespread. I speculate that the reason may be as follows. In both old and new CFE literature the plain symbol $J$ is used, and is said to represent "current density". Most theoretical papers, and most experimental papers relating to single-point-geometry emitters, follow the Stern et al. [3] 1929 convention of using $J$ to represent characteristic local ECD, i.e. the parameter here denoted by $J_C$.

With LAFE technologies, experimental papers determine LAFE-average current density, i.e. the quantity denoted here by $J_M$; but they, too, often use the plain symbol $J$ and the simple name "current density". Thus, conflicting double meanings were created for both the term "current density" and the symbol $J$.

Possibly, experimental researchers, using the second convention and looking for simple theory to support experiments, copied an equation [eq. (2), but without distinguishing subscripts] from a source using the first convention. Although the equation was corrected to depend on macroscopic field, it was not realised that conflicting double meanings existed for *J* as well as for *F* (or *E*). The practice of using this partly-corrected elementary FN-type equation as an equation for $J_M$ may have spread and become accepted because neither researchers nor referees detected ambiguous use of the symbol *J*.

It is slightly difficult to understand why the large discrepancies between the experimental measurements of $J_M$ and the predictions of the eq. (13) have not been noticed and reported more frequently. However, as shown above, it is not entirely straightforward to make valid comparisons of macroscopic and local current densities. Much experimental research has been concerned with values of the characteristic MFEF $\gamma_C$, and with experimental "threshold" voltages (or related threshold macroscopic fields). Perhaps some have noticed a discrepancy in the current densities, have not been able to account for it, have concluded—with some grounds—that it does not affect their own experimental conclusions, and have seen no necessity to comment.

### 6.3 Role of the macroscopic pre-exponential correction factor (MPCF)

The macroscopic pre-exponential correction factor (MPCF) $\lambda_M$ has not previously been discussed as a parameter in its own right. As shown above, when emission is orthodox, $\lambda_M$ can be found using a FN plot and modern SN-barrier theory, and this $\lambda_M$ value also serves as a provisional estimate of the AEE $\alpha_M$. An unexpected conclusion has been that uncertainty in the extracted $\lambda_M$ value is less than that in $\alpha_M$. This shows that $\lambda_M$ should be useful as a parameter for characterising LAFEs—possibly more so than $\alpha_M$. Its role might be better appreciated if it had a more descriptive name, such as *LAFE performance factor*. It will be shown elsewhere that $\lambda_M$ is still a useful parameter when emission is not orthodox.

### 6.4 Proposals for improved practice

Using a defective FN-type equation has not in itself caused serious errors in experimental data interpretation by the LAFE research community, as far as is known. Rather, a major problem is that widespread publication of a defective equation can mislead non-experts, who may mistakenly try to use it to predict LAFE performance. This defective equation thus needs replacing. On the positive side, using an improved equation (and the AEE $\alpha_M$ or MPCF $\lambda_M$) could bring benefits when interpreting CFE data and characterising LAFEs. However, many papers do not report all the experimental information needed to apply the procedure described above. The following proposals aim for improved practice in reporting and analyzing LAFE results.

(1) The meaning of the terms "field" and "current density", and of the symbols $E$ (or $F$) and $J$, as used in a document, should be made explicitly clear by a statement when they are first used. This should indicate whether the field in question is a local barrier field, a macroscopic field as described here, or a gap field, and whether the current density is a local current density or a LAFE-average current density. If either of the terms "field" or "current density" is used with more than one meaning in a given document, then qualifiers should be added to the names, and suffices to the symbols.

(2) For ease of comparison, it is recommended that all current densities should be given in the SI coherent unit A/m$^2$.

(3) When Fowler-Nordheim plots are presented, the units in which the quotient $\{i/V^2\}$ or $\{J_M/F_M^2\}$ was evaluated, before the logarithm was taken, should be indicated (either by labelling the vertical axis, or in the figure caption). The most satisfactory approach is to convert the units to A/V$^2$ before taking the logarithm. For many LAFEs, the result (in order of magnitude terms) may be near $10^{-12}$ A/V$^2$, and the resulting natural logarithm may have a value somewhere near $-30$.

(4) When experimental results are reported as values of $F_M$ and $J_M$, the methods used to derive these parameters from the original current-voltage measurements should be stated, together with the values of the conversion parameters (usually the assumed separation of planar plates, and the assumed LAFE macroscopic area).

(5) When interpreting LAFE experiments, if it is wished to state a FN-type equation, then either eq. (10) or eq. (12) should be used, together with a statement of any approximations involved or subsequently made (such as assuming a SN barrier, or putting $\lambda_C = 1$).

(6) For LAFEs fabricated from non-metals, or that use metal emitters of very small tip radius, it is better to use either eq. (10) or eq. (12) [with $\alpha_M \lambda_C$ written as $\lambda_M$, or vice-versa, if preferred] than to use an elementary FN-type equation. It may be helpful to non-experts to indicate that, (a) strictly, FN-type equations apply only to metal emitters of sufficiently large tip radius, (b) a FN-type equation is being used because of the lack of more appropriate theory, and (c) extracted parameter values are subject to uncertainty..

(7) When a FN-type equation is used for $J_M$ in LAFE papers, then for non-experts it is very unhelpful to include only a theoretical citation to the (slightly flawed) 1928 FN derivation of a basic formula for $J_L$, and to give no indication of subsequent theoretical developments. Ideally, citation(s) should also be given to Ref. [27] (and, maybe, to related references) and to one or more modern accounts of FN theory, such as Refs. [52] or [53], and/or to some specific source of the particular FN-type equation used.

(8) When interpreting FN plot slopes, best practice is to give an equation that contains the generalized slope correction factor $\sigma_M$, and state the value allocated to $\sigma_M$. It is helpful to give the values of both the slope obtained by fitting ($S_M^{\text{expt}}$), and the derived slope characterisation parameter here denoted by $\gamma_C^{*\sigma}$. When reporting slope values, it is helpful to employ the unit "neper (Np)" to indicate that natural (as opposed to "common") logarithms are being used on the vertical axis..

(9) Before interpreting a value $\gamma_C^{*\sigma}$ or $\gamma_C^{*1}$ as a conventional (voltage-independent) field enhancement factor, a test should be applied to establish whether the emission is orthodox.

(10) If an attempt is made to extract "experimental" values for the macroscopic pre-exponential correction factor $\lambda_M$ and/or the area efficiency of emission $\alpha_M$, then the method, equations and approximations used should be indicated, and (ideally) the uncertainties involved should also be indicated.

(11) For technological purposes, $\lambda_M$ is probably a better characterisation parameter than $\alpha_M$.

Two further points needs making. Some countries have strong regulation of their engineering professions, and regard professional engineers as formally responsible for mistakes made in good faith. When it is well established that a defective equation might over-predict current density by a factor of $10^9$ or thereabouts, it is possible to envisage contexts (for example, funding applications and the like) where the accidental use of this defective equation might be regarded by the regulatory body as professional negligence. A purpose of this paper has been to reduce both the misuse of eq. (13), and the possibility of unwanted consequences.

This paper has dealt primarily with the need to use adequate mathematical representations of field emission phenomena, and (in Appendix A) with the mathematics of one method of extracting characterisation parameters. For orthodox emission, the extraction method discussed here works, and the extracted parameters have clear physical interpretations. When emission is not orthodox (which is probably often the case), then the precise interpretation of the FN plot slope may require examination on a case-by-case (or class-by-class) basis, and a slightly different method will be needed to extract and interpret values of $\lambda_M$. The test, in appendix A, for deciding whether a FN plot corresponds to orthodox emission or not, may have an important role in interpreting emission data. Later papers will re-examine some of the published experimental data.

**Appendix A  Mathematics of the parameter extraction procedure**

To estimate $\lambda_M$, and hence $\alpha_M$, one needs a value for $\lambda_C^{-1} J_C$. A convenient approach starts from eq. (4), and assumes a Schottky-Nordheim (SN) barrier. One can define a symbol $J_C^{DD}$ and write

$$\lambda_C^{-1} J_C \approx J_C^{DD} \equiv a\phi^{-1} F_C^2 \exp[-v_F b\phi^{3/2}/F_C], \tag{A1}$$

where $v_F$, as before, is a particular value (see below) of the principal SN barrier function $v$.

The characteristic value $f_C$ of the *scaled barrier field* for a SN barrier of zero-field height $\phi$ is related to the characteristic barrier field $F_C$ by [22]

$$f_C = F_C/F_\phi = c^2 \phi^{-2} F_C, \tag{A2}$$

where $c$ is the Schottky constant [5], $c^{-2} \cong 0.694\,4616$ V nm$^{-2}$ eV$^{-2}$, and $F_\phi$ [$=c^{-2}\phi^2$] is the critical field needed to reduce to zero a SN barrier of zero-field height $\phi$. For the surface location "C", $v_F \equiv v(f_C)$.

Calculations are easiest if eq. (A1) is re-arranged into a *scaled form* (based on $\phi$ and $f_C$ as the independent variables) by defining the $\phi$-dependent parameters $\eta^{SN}$ and $\theta^{SN}$:

$$\eta^{SN} \equiv b\phi^{3/2}/F_\phi = bc^2\phi^{-1/2} \approx 9.836\,238 \times (eV/\phi)^{1/2}, \tag{A3}$$

$$\theta^{SN} \equiv a\phi^{-1} F_\phi^2 = ac^{-4}\phi^3 \approx (7.433\,980 \times 10^{11}\text{ A/m}^2)(\phi/eV)^3. \tag{A4}$$

Thus, eq. (A1) becomes

$$J_C^{DD} = \theta^{SN} f_C^2 \exp[-\eta^{SN} v(f_C)/f_C]. \tag{A5}$$

For illustration, the value $\phi = 4.5$ eV yields $\eta^{SN} \approx 4.64$, $\theta^{SN} \approx 6.77 \times 10^{13}$ A/m$^2$.

Advantages of this form are: (1) it contains a single field-type variable ($f_C$); and (2) one can use a simple good approximation for $v(f_C)$, known to be precise to better than 0.35% over the whole range $0 \leq f_C \leq 1$ [22] and to be better than all simple approximations of equivalent complexity [38]. This yields

$$v(f_C)/f_C \approx 1/f_C - 1 + (\ln f_C)/6 . \tag{A6}$$

For example, for $f_C = 0.25$, formula (A6) yields $v(f_C)/f_C \approx 2.7690$, whereas exact evaluation [22] yields $v(f_C)/f_C \approx 2.7600$.

With these equations, it can be tested whether emission is orthodox, and, if so, then $\lambda_C^{-1} J_C$ can be estimated reliably. The following logical steps are involved.

(1) Using a $J_M$-$F_M$-type FN plot, derive a value for its slope $S_M^{\text{expt}}$, and use eq. (18) to write the corresponding slope characterisation parameter ($\gamma_C^{*\sigma}$) in the form

$$\gamma_C^{*\sigma} = \sigma_M \gamma_C^{*1} = -\sigma_M b \phi^{3/2} / S_M^{\text{expt}} , \tag{A7}$$

where the value of $\sigma_M$ is to be determined, as below.

(2) Decide a value ($F_{Mw}$) of macroscopic field to work with, and note the corresponding experimental value $J_{Mw}$ of macroscopic current density. $F_{Mw}$ should be chosen near the middle of the range of values of $1/F_M$ used when fitting the FN plot.

For orthodox emission, the barrier field $F_{Cw}$ corresponding to macroscopic field $F_{Mw}$ is $F_{Cw} = \gamma_C^{*\sigma} F_{Mw}$. With a SN barrier, the scaled barrier field $f_{Cw}$ corresponding to $F_{Cw}$ can be obtained via eq. (A2). The above steps can be combined into the formula

$$f_{Cw} = -bc^2 \phi^{-1/2} \sigma_M F_{Mw} / S_M^{\text{expt}} . \tag{A8}$$

If one uses the conventional approximation of disregarding all indirect field effects on the FN-plot slope, other than that due to field dependence in the principal SN-barrier function $v$, then $\sigma_M$ in eq. (A8) is given by an appropriate value of the SN-barrier function $s$.

The simplest approximation is to set $\sigma_M = s = 1$, which yields the formula

$$f_{Cw}^{*1} \approx (-9.836\ 238) \cdot (eV/\phi)^{1/2} \cdot F_{Mw}/S_M^{expt} . \tag{A9}$$

A better approximation makes use of a simple good approximation for $s(f)$, namely [22]:

$$s(f) \approx 1 - f/6 . \tag{A10}$$

Substituting eq. (A10) (for $\sigma_M$) into eq. (A8) yields the formulae

$$f_{Cw} \approx f_{Cw}^{*1} / (1 + f_{Cw}^{*1}/6) . \tag{A11}$$

$$s_{Cw} \approx 1 / (1 + f_{Cw}^{*1}/6) . \tag{A12}$$

(3) The above equations can be used to extract "experimentally-derived" values for $f_{Cw}^{*1}$, $f_{Cw}$ and $s_{Cw}$. For bulk metals, one might often expect the derived values of $f_{Cw}^{*1}$ and $f_{Cw}$ to be in the range 0.22 to 0.32, and consequently $s_{Cw}$ to lie in the approximate range 0.947 to 0.963. Thus, an intermediate approximation (for bulk metals) is to use a "typical" $s$-value in this range, such as the value 0.955 used in Section 4.1.

The $f$-value range above (0.22 to 0.32) has been chosen as slightly less than the range (0.20 to 0.34) corresponding to Dyke and Trolan's "safe dc operating conditions" for a tungsten single-point-geometry emitter [37, 49]. If either $f_{Cw}^{*1}$ or $f_{Cw}$ is *well outside* the range 0.22 to 0.32, then this indicates either that a seriously incorrect value has been used for the emitter work-function, or that emission is not orthodox. In most circumstances the first option is not a realistic possibility, and an observed discrepancy means emission is not orthodox. Thus, a useful test can be based on eqns (A9) and (A11).

(4) Equation (A7), with $\sigma_M=1$, can be used to extract a value for $\gamma_C^{*1}$, and, with $\sigma_M=s_{Cw}$, to extract an estimate for $\gamma_C^{*\sigma}$. If emission is orthodox, then $\gamma_C^{*\sigma}$ can be interpreted as a conventional (voltage-independent) field enhancement factors (MFEFs), and $\gamma_C^{*1}$ as an approximation for $\gamma_C^{*\sigma}$.

(5) If emission is orthodox, then eqns (A3) to (A6), with the extracted $f_{Cw}$–value, can be used to estimate a value for $\lambda_C^{-1}J_C$ that corresponds to the value $J_{Mw}$, and hence—via eq. (11)— a value for $\lambda_M$.

(6) The "provisional estimate" for $\alpha_M$, obtained from eq. (11) by taking $\lambda_C=1$, is equal to $\lambda_M$. The uncertainty in $\lambda_C^{-1}$, given in Table 1, can be used to calculate a range of values within which $\alpha_M$ would currently be thought to lie.

It is straightforward to program the complete series of steps using a spreadsheet.

Within the context of the assumptions made, the procedure described above is reasonably self-consistent, as Section 5 shows. However, even for bulk metals, we do not know how valid the conventional assumption is; thus, in the present state of theoretical development, we cannot reliably assess the uncertainty associated with this calculation of $\lambda_C^{-1}J_C$.

When the test detects no discrepancies, the procedure described here *f* above be used as a defined engineering procedure to make formal estimates of $\lambda_C^{-1}J_C$ for non-metals and for metal emitters of small tip radius. However, the results are not physically accurate in these material situations, and this creates additional uncertainty. Again, absence of necessary theory means that it is impossible to reliably estimate this additional uncertainty.

When the test detects a discrepancy, then the interpretations of $\gamma_C^{*\sigma}$ and $\gamma_C^{*1}$ are subject to uncertainty, and reliable estimates of the MPCF $\lambda_M$ and the AEE $\alpha_M$ cannot be made by the method described above. However, a rough "illustrative" value of $\lambda_M$ can be obtained from eq. (11) by taking $\lambda_C^{-1}J_C$ equal to some "illustrative" mid-operating-range value, such as $10^7$ A/m$^2$.

Note that in older literature the SN barrier functions are expressed as functions of the Nordheim parameter $y = +\sqrt{f}$. There are good physical and mathematical reasons [22, 28] for now preferring to use scaled barrier field $f$. In particular, $f$ has a more obvious physical interpretation, and—because it is *linearly* related to barrier field—$f$ is easier to use in practical contexts.

**Table 1**

| Table 1. Illustrative estimates (in January 2012) of specific pre-exponential correction factors in the equation for local emission current density (local ECD) | | | |
|---|---|---|---|
| physical origin of correction factor | symbol individual effect | symbol combined effects | values (multipliers for local ECD) |
| summation over states | $\lambda_{D0}$ | | ~ (0.9 to 1) |
| temperature effects at 300 K | $\lambda_T$ | | ~ 1.1 |
| atomic wave-function effects | $\lambda_E$ | | ~ (0.01 to 10) |
| [all effects on electron supply] | | $\lambda_Z$ (=$\lambda_E \lambda_T \lambda_{D0}$) | ~ (0.01 to 10) |
| tunnelling pre-factor | $P_F$ | | ~ (0.5 to 1) |
| [all effects on local pre-exponential correction factor] | | $\lambda_L$ (=$\lambda_Z P_F$) | ~ (0.005 to 10) |

**Figure Captions**

**Figure 1.** The exact triangular (ET) and Schottky-Nordheim (SN) barriers: (a) as drawn by Fowler and Nordheim (Fig. 3 in Ref. [2]); (b) as calculated exactly, for $H$= 4.5 eV, $F$= 5 V/nm. Note the different proportions of the shaded and unshaded areas inside the triangular barrier, in the two cases.

**Figure 2.** Simulated $J_M$-$F_M$-type Fowler-Nordheim plot, based on eq. (12) and input-parameter values specified in the text.